\documentclass[a4paper,12pt]{article}
\usepackage[dvips]{graphicx}
\usepackage{amssymb,amsmath,epsfig}
\numberwithin{equation}{section}

\begin{document}

\title{Solvable K-essence Cosmologies and  Modified Chaplygin Gas Unified Models of Dark Energy and Dark
Matter}
\author{ M. Sharif$^1$ \thanks{msharif.math@pu.edu.pk}, K.R. Yesmakhanova$^2$,  S. Rani$^1$
\thanks{shamailatoor.math@yahoo.com},
R. Myrzakulov$^2$ \thanks{rmyrzakulov@csufresno.edu}\\
$^1$Department of Mathematics, University of the Punjab,\\
Quaid-e-Azam Campus, Lahore-54590, Pakistan.
\\
$^2$Eurasian International Center for Theoretical Physics,   \\
Eurasian National University, Astana 010008, Kazakhstan}
\date{}

\maketitle
\begin{abstract}
This paper is devoted to the investigation of modified Chaplygin gas
model in the context of solvable k-essence cosmologies. For this
purpose, we construct equations of state parameter of this model for
some particular values of the parameter $n$. The graphical behavior
of these equations are also discussed by using power law form of
potential. The relationship between k-essence and modified Chaplygin
gas model shows viable results in the dark energy scenario. We
conclude that the universe behaves as a cosmological constant,
quintessence phase or phantom phase depending upon $n$.
\end{abstract} \vspace{0.5cm}

\sloppy

\section{Introduction}

The accelerating cosmic expansion of the universe, motivated by the
mysterious dark energy (DE) idea, was noticed in 1998 through
supernovae experiments \cite{1}.  A number of various cosmological
observations claim that this energy fills the space but yet there is
no clue about its identity. The expansion of the universe started to
accelerate when the energy density of matter was overcome by the DE.
The evolution of DE density is governed by an equation of state
parameter, $\omega=p/\rho$. This parameter has raised recently many
interesting features for accelerating expansion of the universe.
Some of them are as follows:
\begin{itemize}
\item Qunitessence region obeys $-1<\omega\leq-1/3$ \cite{y},
\item Cosmological constant is considered to be the simplest case for DE for which
$\omega=-1$,
\item For phantom phase, the EoS parameter follows the domain $\omega<-1$ \cite{x}.
\item Qunitum model inherits both of the properties of
quintessence and phantom phases by crossing the phantom divide line
$\omega=-1$ \cite{w}.
\end{itemize}

Modified Chaplygin Gas (MCG) model \cite{u} is one of the proposed
models to discuss the DE dominated phase for accelerating expansion
of the universe. It is a single fluid model to unify the DE and dark
matter. This model behaves as dark matter when its energy density
evolves as $\rho\propto a^{-3}$ in the early times of the universe
(where the scale factor $a$ is a function of time), whereas the
constant energy density behaves as DE era in late time acceleration.
Thus the MCG model provides unification between dark matter and DE
\cite{v}. The scalar field model is another category to discuss the
DE scenario. Recently, k-essence \cite{z} gains an important role to
explore the dynamics of the universe. It inherits a kinetic term
which deals with late time inflation and early time acceleration
named as k-inflation \cite{2}.

Another DE model, g-essence, is a general essence model reduced in
two models: k-essence (kinetic quintessence) and f-essence
(fermionic k-essence) \cite{3}. Razina et al. \cite{4} explored the
relationship between fermions and scalar fields by taking two
g-essence models. They concluded that g-essence describes the
accelerating expansion of the universe. Tsyba et al. \cite{5}
reconstructed the interaction among Chaplygin gas, Generalized
Chaplygin gas as well as fermions and found EoS parameter for
particular f-essence models. Myrzakulov \cite{6} discussed the
fermionic k-essence to explore accelerating expansion of the
universe. Jamil et al. \cite{7} investigated the behavior of
solvable f-essence cosmology in MCG scenario.

In this paper, we study the purely kinetic k-essence cosmology by
using the phenomenon of MCG model. We obtain a non-linear
differential equation which depends upon the parameter $n$. The EoS
parameter is evaluated by using some particular values of $n$ and is
displayed graphically to investigate the dynamics of the universe.
The power law form of potential function is used to obtain the
kinetic term of purely kinetic k-essence.

The paper is organized as follows. In the next section, we present
some basics of k-essence model and the MCG model. In section
\textbf{3}, the EoS parameter is constructed for different cases and
their graphical representation is discussed. Section \textbf{4}
describes the results for some other forms of solvable k-essence
cosmology. In the last section, we summarize and conclude the
results.

\section{K-essence and MCG Formalism}

Here, we consider the class of models described by the following
action
\begin{equation} \label{1q}
S=\int d^{4}x\sqrt{-g}[R+2K(X, \phi)],
\end{equation}
where $\phi$ is the scalar field, $R$ is the Ricci scalar,
\begin {equation}\nonumber
X=-0.5g^{\mu\nu}\nabla_{\mu}\phi\nabla_{\nu}\phi,
\end{equation}
is the canonical kinetic energy and $g_{\mu\nu}$ is the metric
tensor, $\nabla_{\mu}$ is the covariant derivative associated with
metric $g_{\mu\nu}$. We are interested in flat, homogeneous  and
isotropic FRW universe described by the line element
\begin{equation}\label{1r}
ds^2=-dt^2+a^2(dx^2+dy^2+dz^2),
\end{equation}
where $a(t)$ is the scale factor. The k-essence equations
corresponding to the action \eqref{1q} read as
\begin{eqnarray}\label{1t}
3H^2-\rho &=&0,\\2\dot{H}+3H^2+p&=&0,\\
K_{X}\ddot{\phi}+(\dot{K}_{X}+3HK_{X})\dot{\phi}-K_{\phi}&=&0,\label{1w}\\
\dot{\rho}+3H(\rho+p)&=&0.\label{2w}
\end{eqnarray}
Here dot denotes derivative with respect to cosmic time, the Hubble
rate is $H=\dot{a}/a$, the kinetic term, the energy density  and the
pressure, respectively, are
\begin{equation}\label{1y}
X=0.5\dot{\phi}^2,\quad \rho=2XK_{X}-K,\quad p=K.
\end{equation}
It was shown in \cite{Garriga} that for this model the speed of
propagation of scalar perturbations (speed of sound), $c_s$, is
given by
\begin{equation}\label{1u}
c^2_s=\frac{p_X}{\rho_X}=\frac{p_X}{p_X+2Xp_X}.
\end{equation}
In the case of purely kinetic k-essence, $K=K(X)$, Eq.\eqref{1w}
takes the form
\begin{equation}\nonumber
K_{X}\ddot{\phi}+(\dot{K}_{X}+3HK_{X})\dot{\phi}=0,
\end{equation}
yielding the general solution
\begin{equation}\label{1o}
XK_{X}^2=ka^{-6},
\end{equation}
with $k\geq0$ \cite{Scherrer, Linder}.

The EoS of the MCG DE model is \cite{Benaoum}
\begin{equation}\label{1p}
p=A\rho-\frac{B}{\rho^\alpha},
\end{equation}
where $A$ and $B$ are positive constants and $0\leq\alpha\leq 1.$
Using Eqs.\eqref{2w} and \eqref{1p}, the MCG energy density and
pressure evolve as
\begin{eqnarray}\label{1a}
\rho&=&\left[B(1+A)^{-1}+Ca^{-3(1+\alpha)(1+A)}\right]^{\frac{1}{1+\alpha}},\\
p&=&[ACa^{-3(1+\alpha)(1+A)}-B(1+A)^{-1}]\left[B(1+A)^{-1}
+Ca^{-3(1+\alpha)(1+A)}\right]^{-\frac{\alpha}{1+\alpha}}.
\end{eqnarray}
In this case, the EoS parameter is
\begin{equation}\label{1s}
\omega=\frac{ACa^{-3(1+\alpha)(1+A)}-B(1+A)^{-1}}{B(1+A)^{-1}+Ca^{-3(1+\alpha)(1+A)}}.
\end{equation}
Here we assume the potential in the form of power law \cite{8}
\begin{equation}\label{A}
\phi=\phi_{0}t^\beta.
\end{equation}
Using the above equation, the kinetic term takes the form
\begin{equation}\label{B}
X=0.5\phi_{0}^2\beta^2t^{2(\beta-1)}.
\end{equation}
For the sake of simplicity, we assume $\phi_{0}=1$.

\section{Solvable k-essence cosmology: $K=F(X)$ }

In this section, we would like to consider the case \cite{6} when
\begin{equation}\label{f1}
K=F(X)\end{equation} that is the purely kinetic k-essence case. We
solve the following system
\begin{eqnarray}\label{f2}
\rho&=&2XF_{X}-F,\\ \label{f3} F&=&A\rho-\frac{B}{\rho^\alpha}.
\end{eqnarray}
Solving Eq.\eqref{f2} for $F$, it follows that
\begin{equation}\label{f4}
F=E\sqrt{X}+\frac{\sqrt{X}}{2}\int\frac{\rho}{X^{1.5}}dX,
\end{equation}
where $E$ is an integration constant. Equations \eqref{f3} and
\eqref{f4} give
\begin{equation*}
E\sqrt{X}+\frac{\sqrt{X}}{2}\int\frac{\rho}{X^{1.5}}dX=A\rho-\frac{B}{\rho^\alpha},
\end{equation*}
which has the solution
\begin{equation}\label{f6}
(1+A)\rho^{1+\alpha}-(WX)^{\frac{n(1+\alpha)}{2\alpha}}\rho^{n(1+\alpha)}-B=0,
\end{equation}
where $W$ is a constant and
\begin{equation}\label{f7}
n=\frac{\alpha(1+A)}{A+\alpha(1+A)}.
\end{equation}
From Eq.\eqref{f7}, it follows that $A$ and $\alpha$ are related by
\begin{equation}\label{f8}
A=-\frac{(n-1)\alpha}{(n-1)\alpha+n} \quad or \quad \alpha=-\frac{nA}{(n-1)(1+A)}.
\end{equation}
The search for the analytical solutions of Eq.\eqref{f6} is a hard
job. We would like to find particular solutions for some values of
$n$ to get insights of the model. It is mentioned here that the
values of constants \cite{7,a} in the models are taken in such a way
that we obtain an expanding universe with acceleration.

\subsubsection{Example 1: $n=0$}

First, we consider the particular value $n=0$. It follows from
Eq.\eqref{f7} that this value yields two cases, either $\alpha=0$ or
$A=-1$.

\subsubsection*{Case 1}

Let us consider the case $\alpha=0$. Consequently, Eqs.\eqref{f2}
and \eqref{f3} take the form
\begin{eqnarray}\label{f9}
\rho=2XF_{X}-F,\label{1}\quad F=A\rho-B.
\end{eqnarray}
Solving the above equations, the equation for $\rho$ reads as
\begin{equation*}
(1+A)\rho-(WX)^{\frac{1+A}{2A}}-B=0,
\end{equation*}
which yields
\begin{equation}\label{f11}
\rho=(1+A)^{-1}[B+(WX)^{\frac{1+A}{2A}}].
\end{equation}
The expression for pressure becomes
\begin{equation}\label{f12}
p=(1+A)^{-1}[-B+A(WX)^{\frac{1+A}{2A}}].
\end{equation}
The corresponding EoS parameter is given by
\begin{equation}\label{f13}
\omega=A-\frac{B(1+A)}{B+(WX)^{\frac{1+A}{2A}}}.
\end{equation}
Using Eq.(\ref{B}), the above equation can be written as
\begin{equation}\label{C}
\omega=A-\frac{B(1+A)}{B+(W\beta^{2}t^{2(\beta-1)}/2)^{\frac{1+A}{2A}}}.
\end{equation}
\begin{figure}
\epsfig{file=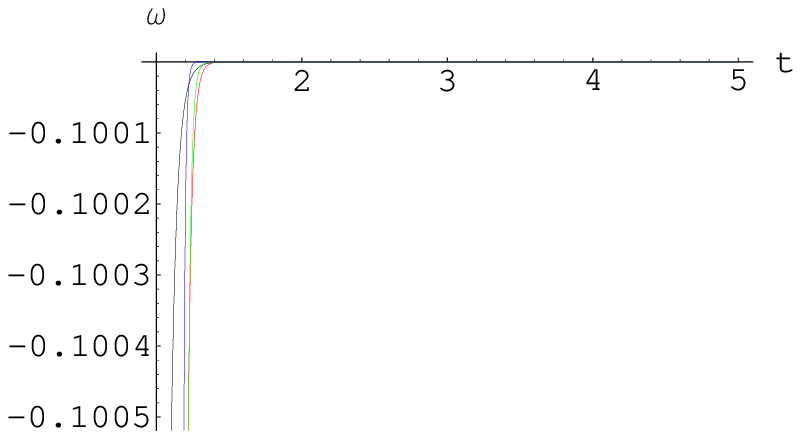, width=0.5\linewidth}\epsfig{file=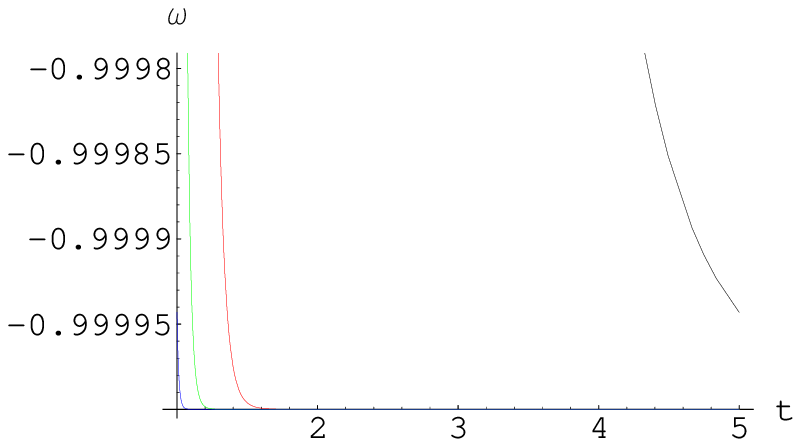,
width=0.5\linewidth}\caption{Plot of $\omega$ versus $t$ with
$A=-0.1,~W=0.2,~B=0.5$. In the left graph, $\beta=-2$ (black),
$\beta=-4$ (red), $\beta=-6$ (green), $\beta=-10$ (blue) and in
right graph $\beta=2$ (black), $\beta=4$ (red), $\beta=6$ (green),
$\beta=10$ (blue).}
\end{figure}
The behavior of EoS parameter $\omega$ in terms of time is shown in
\textbf{Figure 1} which indicates the cosmological evolution. For
$\beta>0$, initially $\omega$ shows the DE phase of the universe
dominated by quintessence. As the value of $\beta$ increases, the
EoS parameter becomes constant at $\omega=-1$ indicating the
universe behavior like cosmological constant. On the other hand, for
$\beta<0$, the EoS parameter possesses negative values and becomes
constant at $-0.1$. It presents a universe which just evolves from
matter dominated universe $(\omega=0)$ to DE energy dominating
universe.

\subsubsection*{Case 2}

Now we consider the case when  $A=-1$. The corresponding equations
\eqref{f2} and \eqref{f3} read as
\begin{equation}\label{f14}
\rho=2XF_{X}-F,\quad F=-\rho-B\rho^{-\alpha},
\end{equation}
which lead to
\begin{equation}\label{f15}
\alpha B\ln\rho-(1+\alpha)^{-1}\rho^{1+\alpha}+\ln (WX)^{\frac{B}{2}}=0.
\end{equation}
Consider some particular solutions  of this equation.

i) If $\alpha=0$, then we have
\begin{equation}\label{f16}
\rho=\ln (WX)^{\frac{B}{2}}
\end{equation}
and for the pressure
\begin{equation}\label{f17}
p=\ln (WX)^{-\frac{B}{2}}-B.
\end{equation}
The corresponding EoS parameter in terms of potential is given by
\begin{equation}\label{f18}
\omega=-1-\frac{2}{\ln [W\beta^{2}t^{2(\beta-1)}/2]}.
\end{equation}
\begin{figure}
\center\epsfig{file=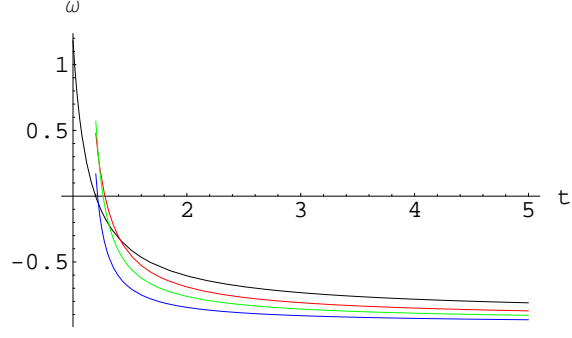, width=0.5\linewidth}\caption{Plot of
$\omega$ versus $t$ with $W=0.2$ and $\beta<0$.}
\end{figure}
The EoS parameter $\omega$ shows the radiation and matter dominated
universe for $t<1.3$ as shown in \textbf{Figure 2}. After this
interval, the universe converges to DE dominated phase lying in the
quintessence region as $-1<\omega \leq-1/3$.

ii) For $\alpha=-1$, the energy density and pressure become
\begin{equation}\label{f20}
\rho=(WX)^{\frac{B}{2(1+B)}},\quad p=-(1+B)(WX)^{\frac{B}{2(1+B)}}.
\end{equation}
The EoS parameter takes the following form
\begin{equation}\label{f22}
\omega=-1-B.
\end{equation}
For this case, the behavior of the universe depends upon the value
of constant $B$. If we take $B=0$, the universe converges to DE
phase dominated by cosmological constant. For $B>0$, the universe
meets the phantom region while $B<0$ shows a quintessence DE
universe.

\subsubsection{Example 2: $n=1$}

If $n=1$, Eq.\eqref{f7} yields $A=0$. This case corresponds to the
generalized Chaplygin gas model \cite{Linder} for which the
expressions of $\rho$ and $p$ are as follows
\begin{eqnarray}\label{a}
\rho=B^{\frac{1}{1+\alpha}}[1-(WX)^\frac{1+\alpha}{2\alpha}]^{\frac{-1}{1+\alpha}},\label{b}\\
p=-B^{\frac{1}{1+\alpha}}[1-(WX)^\frac{1+\alpha}{2\alpha}]^{\frac{\alpha}{1+\alpha}}.
\end{eqnarray}
The corresponding EoS parameter is given by
\begin{equation}\label{c}
\omega=-1+(WX)^\frac{1+\alpha}{2\alpha}.
\end{equation}
Inserting the value of $X$, Eq.(\ref{c}) becomes
\begin{equation}\label{D}
\omega=-1+\left(\frac{W\beta^{2}t^{2(\beta-1)}}{2}\right)^\frac{1+\alpha}{2\alpha}.
\end{equation}
\textbf{Figure 3} shows the graphical behavior of the EoS parameter
with respect to evolving time. The universe acquires quintessence
region initially. As the time elapses, it becomes constant at
$\omega=-1$ showing a universe dominated by the cosmological
constant.
\begin{figure}
\center\epsfig{file=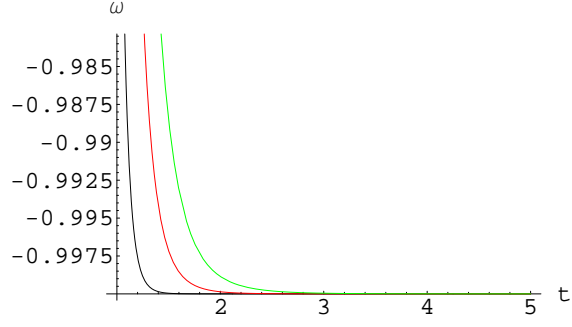, width=0.5\linewidth}\caption{Plot of
$\omega$ versus $t$ with $W=0.2,~\beta=-2$ and $\alpha=0.2$ (black),
$\alpha=0.4$ (red), $\alpha=0.6$ (green).}
\end{figure}

\subsubsection{Example 3: $n=2$}

Now we consider the case when $n=2$. Here $A$ and $\alpha$ are
related by the formula
\begin{equation}\label{f23}
A=-\frac{\alpha}{\alpha+2} \quad or\quad
\alpha=-\frac{2A}{1+A}.
\end{equation}
The equation for $\rho$ \eqref{f6} takes the form
\begin{equation}\label{f24}
(1+A)\rho^{1+\alpha}-(WX)^{\frac{1+\alpha}{\alpha}}\rho^{2(1+\alpha)}-B=0,
\end{equation}
which has the solution
\begin{equation}\label{f25}
\rho=(WX)^{-\frac{1}{\alpha}}\left[\frac{1+A}{2}\mp\frac{1}{2}\sqrt{(1+A)^2-4B(WX)^{\frac{1+\alpha}
{\alpha}}}\right]^{\frac{1}{1+\alpha}}
\end{equation}
or
\begin{equation}\label{f25}
\rho=(WX)^{-\frac{1}{\alpha}}\left\{\frac{1}{2+\alpha}
\left[1\mp\sqrt{1-B(2+\alpha)^2(WX)^{\frac{1+\alpha}
{\alpha}}}\right]\right\}^{\frac{1}{1+\alpha}}
\end{equation}
The pressure is given by
$$p=-\frac{\alpha(WX)^{-\frac{1}{\alpha}}}{\alpha+2}
\left\{\frac{1}{2+\alpha}\left[1\mp\sqrt{1-B(2+\alpha)
^2(WX)^{\frac{1+\alpha}{\alpha}}}\right]\right\}^{\frac{1}{1+\alpha}}$$
\begin{equation}\label{f26}
-BWX\left\{\frac{1}{2+\alpha}
\left[1\mp\sqrt{1-B(2+\alpha)^2(WX)^{\frac{1+\alpha}
{\alpha}}}\right]\right\} ^{-\frac{\alpha}{1+\alpha}}.
\end{equation}
The corresponding EoS parameter is
\begin{equation}\label{f27}
\omega=-\frac{\alpha}{\alpha+2}-B(WX)^{\frac{1+\alpha}{\alpha}}\left\{\frac{1}{1+\alpha}\left[1\mp\sqrt{1-
B(2+\alpha)^2(WX)^{\frac{1+\alpha}{\alpha}}}\right]\right\}^{-1}.
\end{equation}
\begin{figure}
\center\epsfig{file=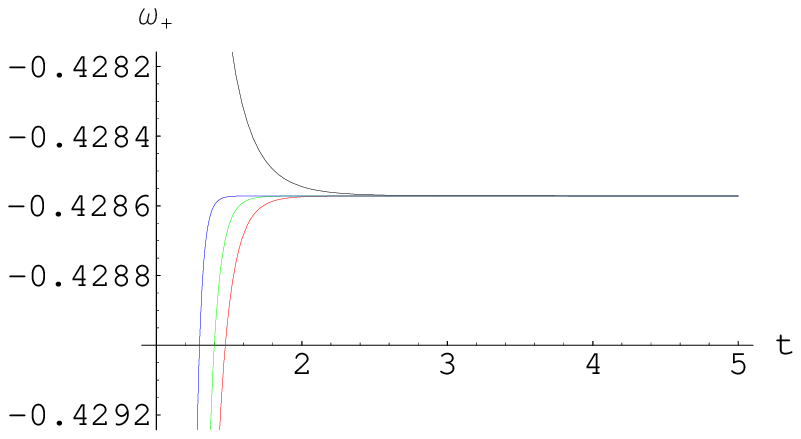, width=0.5\linewidth}\caption{Plot of
$\omega$ versus $t$ with $W=0.2,~B=-0.1,~\alpha=1.5$ and $\beta<0$.}
\end{figure}
Inserting the value of potential and after simplifying the positive
root, it yields
\begin{equation}\label{E}
\omega_+=-\frac{\alpha}{\alpha+2}-\frac{2B(1+\alpha)
(W\beta^{2}t^{2(\beta-1)})^{\frac{1+\alpha}{\alpha}}}{2^{\frac{1+3\alpha}{\alpha}}
-B(\alpha+2)^{2}(W\beta^{2}t^{2(\beta-1)})^{\frac{1+\alpha}{\alpha}}}.
\end{equation}
For this case, the EoS parameter results a universe having dynamics
of quintessence of DE scenario shown in \textbf{Figure 4}. It
becomes constant at $\omega_{+}=-0.4285$ with the passage of time.
The negative root of Eq.(\ref{f27}) does not give any fruitful
result.

\subsubsection{Example 4: $n=0.5$}

If $n=0.5$ then $A$ and $\alpha$ satisfy the relation
\begin{equation}\label{f28}
A=-\frac{\alpha}{\alpha-1} \quad or\quad
\alpha=\frac{A}{1+A}.
\end{equation}
The equation \eqref{f6} becomes
\begin{equation}\label{f29}
(1+A)\rho^{1+\alpha}-(WX)^{\frac{1+\alpha}{4\alpha}}\rho^{0.5(1+\alpha)}-B=0.
\end{equation}
The solution of this equation reads as
\begin{equation}\label{f30}
\rho=[2(1+A)]^{-\frac{2}{1+\alpha}}\left[(WX)^{\frac{1+\alpha}{4\alpha}}\pm\sqrt{(WX)^{\frac{1+\alpha}{2\alpha}}
+4B(1+A)}\right]^{\frac{2}{1+\alpha}},
\end{equation}
or
\begin{equation}\label{f30}
\rho=\left(\frac{2}{1-\alpha}\right)^{-\frac{2}{1+\alpha}}
\left[(WX)^{\frac{1+\alpha}{4\alpha}}\pm\sqrt{(WX)^{\frac{1+\alpha}{2\alpha}}
+\frac{4B}{1-\alpha}}\right]^{\frac{2}{1+\alpha}}.
\end{equation}
The pressure is given by
$$p=\frac{\alpha}{1-\alpha}\left(\frac{2}{1-\alpha}\right)^{\frac{-2}{1+\alpha}}
\left[(WX)^{\frac{1+\alpha}{4\alpha}}\pm\sqrt{(WX)^{\frac{1+\alpha}{2\alpha}}+
\frac{4B}{1-\alpha}}\right]^{\frac{2}{1+\alpha}}$$
\begin{equation}\label{f31}
-B\left(\frac{2}{1-\alpha}\right)^{\frac{2\alpha}{1+\alpha}}
\left[(WX)^{\frac{1+\alpha}{4\alpha}}\pm\sqrt{(WX)^{\frac{1+\alpha}{2\alpha}}
+\frac{4B}{1-\alpha}}\right]^{-\frac{2\alpha}{1+\alpha}}.
\end{equation}
The corresponding EoS parameter is
\begin{equation}\label{f32}
\omega=-\frac{\alpha}{\alpha-1}-
4B\left(\frac{1}{1-\alpha}\right)^{2}\left[(WX)^{\frac{1+\alpha}{4\alpha}}\pm
\sqrt{(WX)^{\frac{1+\alpha}{2\alpha}}+\frac{4B}{1-\alpha}}\right]^{-2}.
\end{equation}
Using Eq.\eqref{B} and after some manipulation, the EoS parameter
takes the form
\begin{equation}\label{G}
\omega_+=-\frac{\alpha}{\alpha-1}-\frac{B
(2W\beta^{2}t^{2(\beta-1)})^{\frac{1+\alpha}{2\alpha}}}
{[(\alpha-1)(W\beta^{2}t^{2(\beta-1)})^{\frac{1+\alpha}{2\alpha}}-2^{\frac{1+\alpha}{2\alpha}}B]^2}.
\end{equation}
\begin{figure}
\center\epsfig{file=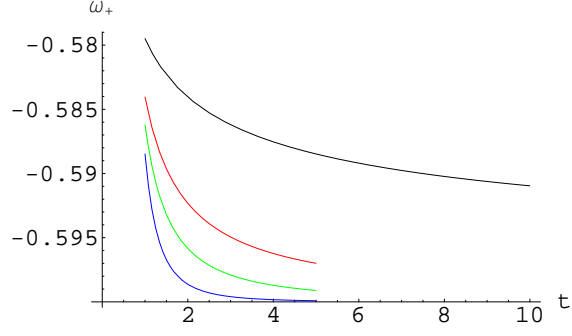, width=0.5\linewidth}\caption{Plot of
polytropic k-essence $\omega_+$ versus $t$ with
$W=0.2,~B=-0.1,~\alpha=-1.5$ and $\beta>0$.}
\end{figure}
\textbf{Figure 5} shows the same behavior of EoS parameter as
discussed for $n=2$. However, this remains in the region,
$-1<\omega_{+}<-1/3$, at different points. In this case, the EoS
parameter becomes polytropic k-essence EoS parameter as $\alpha<0$.
It is concluded that this kind of parameter shows the accelerating
expansion of the universe.

\subsubsection{Example 5: $n=-1$}

For this example, $A$ and $\alpha$ satisfy the relation
\begin{equation}\label{f33}
A=-\frac{2\alpha}{2\alpha+1} \quad or\quad
\alpha=-\frac{A}{2(1+A)}.
\end{equation}
Consequently, Eq.\eqref{f6} becomes
\begin{equation}\label{f34}
(1+A)\rho^{1+\alpha}-(WX)^{-\frac{1+\alpha}{2\alpha}}\rho^{-(1+\alpha)}-B=0,
\end{equation}
which yields the solution
\begin{equation}\label{f35a}
\rho=\left[\frac{B\pm\sqrt{B^2+4(1+A)(WX)^{-\frac{1+\alpha}
{2\alpha}}}}{2(1+A)}\right]^{\frac{1}{1+\alpha}}.
\end{equation}
or
\begin{equation}\label{f35b}
\rho=\left[\frac{B(2\alpha+1)}{2}\pm\frac{2\alpha+1}{2}\sqrt{B^2+\frac{4}{2\alpha+1}(WX)^{-\frac{1+\alpha}
{2\alpha}}}\right]^{\frac{1}{1+\alpha}}.
\end{equation}
The pressure is
$$p=-\frac{2\alpha}{2\alpha+1}\left[\frac{B(2\alpha+1)}{2}\pm\frac{2\alpha+1}{2}\sqrt{B^2
+\frac{4}{2\alpha+1}(WX)^{-\frac{1+\alpha}
{2\alpha}}}\right]^{\frac{1}{1+\alpha}}$$
\begin{equation}\label{f36}
-B\left[\frac{B(2\alpha+1)}{2}\pm\frac{2\alpha+1}{2}\sqrt{B^2+\frac{4}{2\alpha+1}(WX)^{-\frac{1+\alpha}
{2\alpha}}}\right]^{-\frac{\alpha}{1+\alpha}}.
\end{equation}
The corresponding EoS parameter yields
\begin{equation}\label{f37}
\omega=-\frac{2\alpha}{2\alpha+1}-B\left[\frac{B(2\alpha+1)}{2}\pm\frac{2\alpha+1}{2}\sqrt{B^2
+\frac{4}{2\alpha+1}(WX)^{-\frac{1+\alpha}{2\alpha}}}\right]^{-1}.
\end{equation}
Inserting the value of kinetic term $X$, the positive and negative
roots of EoS parameter $\omega$ become
\begin{eqnarray}\label{H}
\omega_+&=&-\frac{2\alpha}{2\alpha+1}
-\frac{B^{2}(W\beta^{2}t^{2(\beta-1)})^{\frac{1+\alpha}{2\alpha}}}{B^{2}(2\alpha+1)
(W\beta^{2}t^{2(\beta-1)})^{\frac{1+\alpha}{2\alpha}}
+2^{\frac{1+\alpha}{2\alpha}}},\label{I}\\
\omega_-&=&-\frac{2\alpha}{2\alpha+1}+{B^{2}
\left(\frac{W\beta^{2}t^{2(\beta-1)}}{2}\right)^{\frac{1+\alpha}{2\alpha}}}.
\end{eqnarray}
\begin{figure}
\epsfig{file=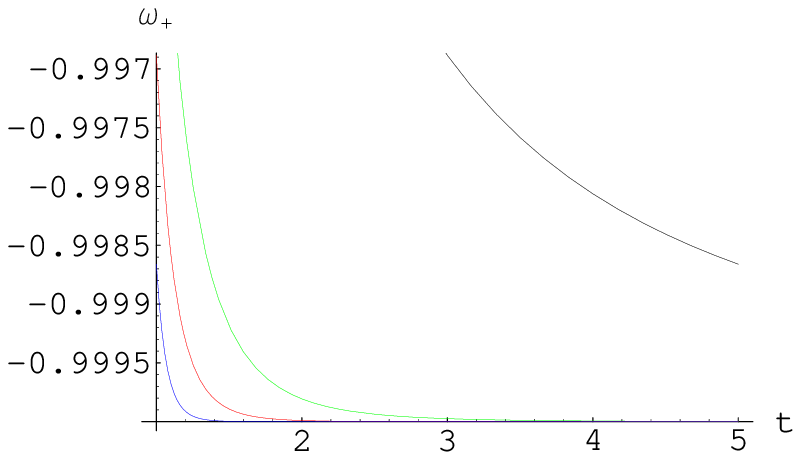, width=0.5\linewidth}\epsfig{file=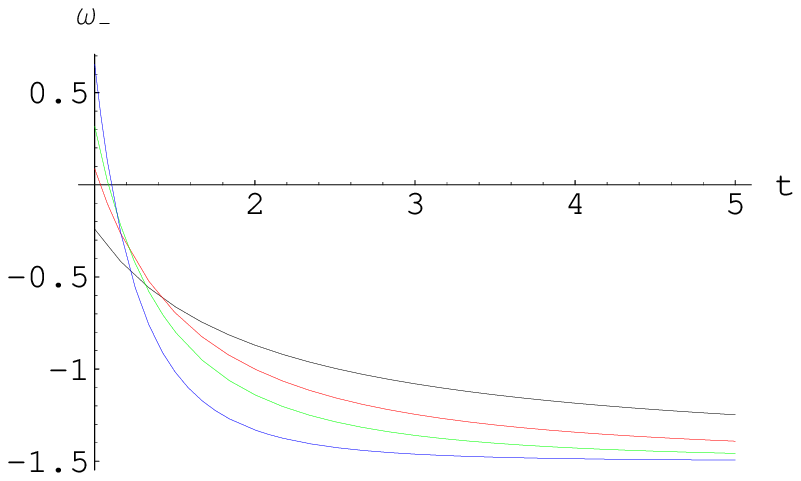,
width=0.5\linewidth}\caption{Plot of $\omega$ versus $t$ with $W=2$
and $B=-1$. In the left graph, $\beta>0,~\alpha=1.5$ and in right
graph $\beta<0,~\alpha=-1.5$.}
\end{figure}

The left graph in \textbf{Figure 6} shows the cosmological evolution
of EoS parameter $\omega_+$ with respect to time. It represents a
quintessence region of the universe for higher values of $\beta$
while for smaller values, it becomes constant and behaves like
cosmological constant. In the right graph, polytropic k-essence EoS
parameter is plotted. It represents initially a matter dominated
universe for $t<1.2$ for smaller values of $\beta$. Then $\omega_-$
crosses the phantom divide line from quintessence to phantom region
and becomes constant in the phantom region $(\omega_-<-1)$.

\section{Other Forms of k-essence Cosmology}

There are two other possibilities of solvable k-essence cosmology
\cite{6} such as:
\begin{itemize}
\item General k-essence cosmology: $K(X,\phi)=V(\phi)F(X)$
\item Quintessence k-essence cosmology: $K(X,\phi)=V(\phi)+F(X)$
\end{itemize}
These forms yield the same equations (\ref{f6}) and (\ref{f7}) for
$\rho$ and  for parameter $n$. It is mentioned here that the above
forms show the same behavior as in the case of purely kinetic
k-essence cosmology.

Finally, we would like to mention the other class of solvable
k-essence models namely integrable k-essence models \cite{Fam1}. As
an example, let us consider the following so-called E$_{IIE}$-model
\begin{equation}
K_{XX}=2K^3+XK+\alpha(\phi),
\end{equation}
where $\alpha=\alpha(\phi)$. It is nothing but the
Painlev$\acute{e}$ II (P-II) equation which is known as integrable.
Let the solution of this equation we define as $K=K(X;\alpha)$ and
here we assume that $\alpha=constant$. Then the equation (4.1)  has
the following particular solutions (see \cite{Fam1} and references
therein)
\begin{eqnarray}
K&\equiv &K(X;1.5)=\psi-(2\psi^2+X)^{-1},\\
K&\equiv &K(X;1)=-\frac{1}{X}, \\
K&\equiv &K(X;2)=\frac{1}{X}-\frac{3X^2}{X^3+4},\\
K&\equiv& K(X;3)=\frac{3X^2}{X^3+4}-\frac{6X^2(X^3+10)}{X^6+20X^3-80},\\
K&\equiv& K(X;4)=-\frac{1}{X}+\frac{6X^2(X^3+10)}{X^6+20X^3-80}-\frac{9X^5(X^3+40)}{X^9+60X^6+11200},\\
K&\equiv& K(X;0.5\epsilon)=-\epsilon\psi
\end{eqnarray}
and so on. Here
\begin{equation}
\psi=(\ln{\chi})_{X}, \quad  \chi(X,\phi)=C_1(\phi)Ai(-2^{-1/3}X)+C_2(\phi)Bi{(-2^{-1/3}X)},
\end{equation}
where $C_i=C_i(\phi)$ and $Ai(x), Bi(x)$ are Airy functions. Each of
these solutions determine different type of cosmology. For example,
the solution (4.4) corresponds to the EoS:
\begin{eqnarray}
p&=&\frac{2(2-X^3)}{X(X^3+4)},\\
\rho&=&\frac{6(X^6-10X^3-8)}{X(X^3+4)^2}.
\end{eqnarray}
Hence we get the EoS parameter
\begin{equation}
\omega=-\frac{1}{3}-\frac{4X^3}{X^6-10X^3-8}.
\end{equation}
Inserting the value of kinetic term (2.15), this equation takes the
form
\begin{equation}
\omega=-\frac{1}{3}-\frac{32\phi_0^6\beta^6t^{6(\beta-1)}}
{\phi_0^{12}\beta^{12}t^{12(\beta-1)}-80\phi_0^6\beta^6t^{6(\beta-1)}-512}.
\end{equation}
If $\beta=1=\phi_0$, then $\omega\sim-0.27$. This shows that the
solution (4.4) represents a universe which is nearly to meet the
quintessence phase of the accelerating universe. Similarly, we can
calculate the EoS and its parameter for the other solutions from the
system (4.2)-(4.7) and as well as for the other integrable k-essence
models from \cite{Fam1}. Finally we note that it is interesting to extend results of this work to the periodical and  quasi-periodical gas/fluid models (see e.g. Ref. \cite{Fam2}) as well as to the knot universe models \cite{Fam3}.

\section{Conclusion}

It has always been interesting to explore models to study the
accelerating expansion of the universe. This paper is devoted to the
study of some solvable k-essence cosmologies using MCG. In
particular, we have discussed the purely kinetic k-essence model by
taking the energy density and pressure of MCG model. We have found a
nonlinear differential equation for the energy density which depends
upon the parameter $n$. We have constructed EoS parameter for
different values of $n$ by using power law form of potential. Also,
the values of constants involved in the model such as,
$A,~B,~\alpha$ and $\beta$ are constrained in such a way that they
give the compatible behavior of our expanding universe with
acceleration. The graphical results can be summarized as follows:
\begin{itemize}
\item  Example $1$ is taken as $n=0$ which further leads to two
cases. The case $\alpha=0$ gives accelerating expansion of the
universe dominated by cosmological constant for $\beta>0$, while
$\beta<0$ indicates a universe when DE is dominated over the matter
to accelerate the universe. The second case, $A=-1$, indicaress the
dynamics of the universe underlying in quintessence phase.
\item The second example leads to the generalized Chaplygin gas model for
$n=1$. For this value of parameter, initially our universe shows
quintessence behavior of DE dominated era. After that, it converges
to $\omega=-1$ representing cosmological constant.
\item  For $n=2$ and $\beta<0$, the EoS parameter shows quintessence behavior of the universe.
\item  The case $n=0.5$, using k-essence polytropic EoS parameter,
gives the same dynamics of the universe as discussed for the
parameter $n=2$.
\item  For $n=-1,~\omega_+$ leads to a cosmological
constant dominated universe for smaller values of $\beta$ whereas
the k-essence polytropic EoS parameter $\omega_-$ represents a
quintom behavior and becomes constant in the phantom region.
\end{itemize}

It is concluded that this relationship provides consistent results
about accelerating expansion of the universe \cite{7}. We would like
to mention here that the same behavior is observed for the general
k-essence and quintessence cosmologies as discussed for puerly
kinetic k-essence cosmology by using MCG model. Finally, we have
given some integrable k-essence cosmologies $K(X,\alpha)$. These
models show different phases of the accelerating universe depending
upon its parameters.


\begin{thebibliography}{99}

\bibitem{1} Perlmutter, S. et al.: Astrophys. J. \textbf{517}(1999)565.

\bibitem{y} Carroll, S.M.: Phys .Rev. Lett. \textbf{81}(1998)3067.

\bibitem{x} Yadav, A.K.: Astrophys. Space Sci.
\textbf{335}(2011)565.

\bibitem{w} Khatua, P.B., Chakraborty, S. and Debnath, U.: arXiv:1105.3393.

\bibitem{u} Bento, M.C., Bertolami, O. and Sen, A.A.: Phys. Rev.
\textbf{D66}(2002)043507.

\bibitem{v} Matarrese, S. et al.: \textit{Dark Matter and Dark Energy}
(Springer, 2010).

\bibitem{z} Mukhanov, V., Armendariz-Picon, C. and Steinhardt, P.J.: Phys.
Rev. Lett. \textbf{85}(2010)4438.

\bibitem{2} Piazza, F. and Tsujikawa, S.: JCAP \textbf{07}(2004)004.

\bibitem{3} Yerzhanov, K.K. et al.: arXiv:1012.3051.

\bibitem{4} Razina, O. et al.: Eur. Phys. J. Plus 85(2011)126.

\bibitem{5} Tsyba, P. et al.: arXiv:1103.5918.

\bibitem{6} Myrzakulov, R.: arXiv:1011.4337.

\bibitem{7} Jamil, M. et al.: Astrophys. Space Sci. \textbf{336}(2011)325.

\bibitem{Garriga} Garriga, J. and Mukhanov, V.F.: Phys. Lett. {\bf B458}(1999)219.

\bibitem{Scherrer} Scherrer, R.: Phys. Rev. Lett.
{\bf93}(2004)011301.

\bibitem{Linder} De Putter, R. and Linder E.V.: Astropart. Phys. {\bf28}(2007)
263.

\bibitem{Benaoum} Benaoum H.B.: arXiv:0205140v1.

\bibitem{8} Jamil, M., Hussain, I. and Momeni, D.: Eur. Phys. J.
Plus \textbf{126}(2011)80.

\bibitem{a} Jamil, M., Rahaman, F. and Kalam, M.: Eur. Phys. J.
\textbf{C60}(2009)149.

\bibitem{Fam1}  Myrzakul Sh.R.,  Esmakhanova K.R.,  Myrzakulov K.R.,  Nugmanova G.N., Myrzakulov R.: arXiv:1105.2771.
\bibitem{Fam2}   Bamba K.,  Yesmakhanova K.,  Yerzhanov K., Myrzakulov R.: arXiv:1203.3401; \\ Bamba K.,  Debnath U.,  Yesmakhanova K.,  Tsyba P.,  Nugmanova G.,  Myrzakulov R.: arXiv:1203.4226; \\ Bamba K.,  Razina O.,  Yerzhanov K., Myrzakulov R.: arXiv:1203.2804. 
\bibitem{Fam3}  Myrzakulov R.: arXiv:1008.4486;\\ Yesmakhanova K.,  Myrzakulov Y., Nugmanova G.,  Myrzakulov R.: International Journal of Theoretical Physics, {\bf 51} (2012) 1204-1210, arXiv:1102.4456; \\Myrzakulov R.:arXiv:1205.5266;\\Myrzakulov R.:arXiv:1207.1039;\\ Myrzakulov R. et al.: arXiv:1201.4360;\\ Myrzakulov R.: arXiv:1204.1093. 
\end{thebibliography}
\end{document}